# Contact resistance at planar metal contacts on bilayer graphene and effects of molecular insertion layers

**Ryo Nouchi**[1]

Nanoscience and Nanotechnology Research Center, Osaka Prefecture University, Sakai, Osaka 599-8570, Japan

______________________________

[1]Author to whom correspondence should be addressed.
E-mail: r-nouchi@21c.osakafu-u.ac.jp

**Abstract**

The possible origins of metal–bilayer graphene (BLG) contact resistance are investigated by taking into consideration the bandgap formed by interfacial charge transfer at the metal contacts. Our results show that a charge injection barrier (Schottky barrier) does not contribute to the contact resistance because the BLG under the contacts is always degenerately doped. We also showed that the contact-doping-induced increase in the density of states (DOS) of BLG under the metal contacts decreases the contact resistance owing to enhanced charge carrier tunnelling at the contacts. The contact doping can be enhanced by inserting molecular dopant layers into the metal contacts. However, carrier tunnelling through the insertion layer increases the contact resistance, and thus, alternative device structures should be employed. Finally, we showed that the inter-band transport by variable range hopping via in-gap states is the largest contributor to contact resistance when the carrier type of the gated channel is opposite to the contact doping carrier type. This indicates that the strategy of contact resistance reduction by the contact-doping-induced increase in the DOS is effective only for a single channel transport branch (*n*- or *p*-type) depending on the contact doping carrier type.

# 1. Introduction

One-atom-thick sheets obtained from graphite, a layered material, are called graphene. Its ultrahigh charge carrier mobility makes it a promising material for future high-speed electronics applications [1,2]. However, single-layer graphene possesses no bandgap, making its applications in field-effect transistors (FETs) in digital logic circuits difficult. For this purpose, AB-stacked bilayer graphene (BLG) has attracted much attention because a finite bandgap can be introduced by applying a potential difference between the two constituent graphene layers [3–7]. The carrier mobility of BLG is very high and close to that of its single-layer counterpart [8,9]; hence, BLG is now considered as a promising material for digital applications.

The biggest difference between BLG and other semiconductors that have been conventionally used for digital applications is its tuneable bandgap, which can be changed by changing the degree of potential difference introduced to the constituent graphene layers [3]. An inter-layer potential difference can be introduced by an out-of-plane electric field [4,5] and interfacial charge transfer (CT) from surface adsorbates [6,7]. In an FET structure schematically shown in Fig. 1(a), both of these two sources can be found, and the source for the former is gate voltage. While a double gate structure has been widely used for introducing a sizeable bandgap maintaining the controllability of charge carrier density in BLG [4,5,10–12], the single gate structure displayed in Fig. 1(a) also generates a vertical electric field. The source of the latter is the source/drain electrode contacts. At these metal contacts, interfacial CT should naturally occur because the Fermi level of the electrodes and the underlying BLG layer should be aligned. While the CT from surface adsorbates with a high ionization potential or low electron affinity has been well documented [6,7,13–15], the CT from conventional electrode metals has been investigated less [16,17].

The electrode contacts are the origin of a parasitic resistance called contact resistance, which dominates the total device resistance of very small FETs [18], and should be eliminated as much as possible. The CT at the contacts is known to greatly affect the magnitude of contact resistance in graphene-related systems. In the case of single-layer graphene, which is a zero-bandgap material, the contact resistance is mainly determined by tunnel resistance at the metal/graphene interface. The tunnel resistance is inversely proportional to the density of states (DOS) at the Fermi level of the two systems at both sides of the tunnel junction. The DOS of single-layer graphene increases as the Fermi level is shifted away from its charge neutrality point (the so-called Dirac point) [19]. Thus, the CT from the electrode metal increases the DOS of the underlying graphene, leading to a lower contact resistance [20,21]. In the case of BLG, the situation is rather complicated because of its tuneable bandgap. The DOS-related tunnel resistance is as important as for single-layer graphene, and decreases if the amount of CT increases. However, the resistance originating from the inter-band transport across the bandgap should also be taken into consideration, as shown in Fig. 1(b) [17]. The CT from the metal contacts generates a bandgap in the underlying BLG. Thus, if the amount of CT increases, the bandgap becomes wider, and this should increase the contact resistance. Thus, in the case of BLG, two competing effects of the CT should be carefully examined.

In the present study, the interfacial CT at the metal contacts on BLG is discussed in detail. Possible sources of the contact resistance are independently examined, and inter-band transport is shown to be likely the most dominant contribution. In addition, a molecular interfacial layer with low electron affinity is tested as a possible method that can reduce the contact resistance via increased interfacial CT.

## 2. Sample preparation

BLG flakes were mechanically exfoliated onto 300-nm-thick thermally oxidized silicon on a heavily *p*-doped silicon substrate using adhesive tape. The number of exfoliated graphene layers was confirmed to be two by optical microscope contrast and Raman spectroscopy. Electron beam lithography was performed to pattern electrodes, followed by the deposition of Cr (as an adhesion layer) and Au electrode (1 and 50 nm thick, respectively) using vacuum deposition (~$10^{-4}$ Pa) and a lift-off process. FETs with interfacial molecular layers were fabricated by forming a molecular layer prior to the metal deposition. Multiple electrodes with different spacing were fabricated on a single BLG flake in order to determine the contact resistance using the transfer length method (TLM) [22]. Electrical characterizations of the fabricated FETs were performed using a source meter (Keithley Instruments Inc., 2636A) under dark conditions.

The molecular interfacial layer was formed by immersing the substrate into a 1 mM aqueous solution (including 1 wt% of sodium dodecyl sulfate) of 4-nitrobenzene diazonium tetrafluoroborate (NDT). The immersion was conducted after electron-beam patterning of the electron-beam resist. After immersion into the solution at 40 °C for 4 h (with stirring at 120 rpm), the substrates were thoroughly rinsed with distilled water. NDT is known as a hole dopant for graphene because of the presence of the electron-withdrawing nitro group [23], but can break the $sp^2$ network by rehybridising to $sp^3$ along with covalent bond formation with the aryl group of NDT [24–26]. While the interaction between the first monolayer of NDT and BLG is strong because of the covalent bond, the interaction between the second (and more) NDT monolayer and the underlying NDT layers should be weak because of its physisorbed nature. Thus, only the first monolayer that makes a strong covalent bond is likely to remain after the rinse process. The substrate was then introduced into a vacuum deposition chamber immediately after the immersion in order to minimize the adsorption of hydrocarbon pollutants from the air.

## 3. Results

Figure 2(a) shows the transfer characteristics of an as-fabricated BLG FET without the NDT interfacial layer, measured in air at 300 K. The channel length and width of the measured FET are 1.0 and 2.6 µm, respectively. The gate voltage that corresponds to the charge neutrality point, $V_{NP}$, of BLG is at 33 V, which indicates a hole-doped channel. Possible origins of the hole doping include presence of adsorbates from air such as oxygen and water molecules [27,28] and source/drain metal contacts [29,30], with the latter considered important in the present study. The amount and type of carriers doped from the metal contact can be determined by examining the channel-length dependence of $V_{NP}$ [31], shown in Fig. 2(b), along with the corresponding amount of the Fermi level shift from the charge neutrality point, $\Delta E_F = \hbar^2 \pi \varepsilon_0 \varepsilon_r V_{NP}/(2dqm^*)$, where $\hbar$ is the Dirac constant, $\varepsilon_0$ is the permittivity of vacuum, $\varepsilon_r$ is the relative permittivity of the gate insulator (3.9), $d$ is the thickness of the gate insulator (300 nm), $q$ is the unit electronic charge, and $m^* = 0.033m_0$ ($m_0$ is the bare electron mass) is the effective mass of the carriers in BLG [32]. By extrapolating to zero channel length, the value of $\Delta E_F$ at the contacts was estimated to be 0.14 eV. This corresponds to a hole concentration of $3.8 \times 10^{12}$ cm$^{-2}$, and thus, higher (valence) bands can be excluded here [33].

Figures 2(c) and 2(d) show the transfer characteristics and channel-length-dependent $V_{NP}$ of BLG FETs with the NDT interfacial layer, respectively. The channel length and width of the FET in Fig. 2(c) are 1.0 and 1.3 µm, respectively. By extrapolating to the zero channel length in Fig. 2(d), the value of $\Delta E_F$ at the contacts was estimated to be 0.24 eV. This indicates that the insertion of the NDT layer indeed increased the amount of CT underneath the metal contacts. The $\Delta E_F$ value corresponds to a hole concentration of $6.6 \times 10^{12}$ cm$^{-2}$, and the higher (valence) bands [33] can be excluded in this case.

From the channel-length dependence of the transfer characteristics, contact resistances, $R_C$, can be determined by means of the TLM. As schematically shown in Fig. 3(a), $R_C$ can be determined from the y-intercept of a linear fit to the plots of the total device resistance vs. the channel length. The calculations should be done with respect to the relative gate voltage, $V_G$, from $V_{NP}$, which is necessary to avoid artefacts in the $R_C$–$V_G$ characteristics [34]. The determined $R_C$ values normalized with the channel width, $W$, are shown in Fig. 3(b). The $R_C$ values in the positive gate region are higher than those in the negative gate region. This feature is a signature of hole doping underneath the contacts [17].

The $R_C$ at metal–BLG contacts is largely affected by the inter-band transport across the bandgap as shown in Fig. 1(b), which can be detected as superlinearity in the high-field current–voltage characteristics [17]. Figure 4(a) shows the output characteristics of a BLG FET without the NDT layer measured at 82 K after annealing at 500 K for 3 h in a vacuum of $3 \times 10^{-3}$ Pa. This FET is the same as the 0.5-µm channel in Fig. 2(b), and the channel width is 3.0 µm. The $V_G$ was swept from $V_{NP} - 40$ to $V_{NP} + 40$ V in 5 V steps, and superlinear curves are clearly observed. To parameterize the nonlinearity, the differential conductances ($dI_D/dV_D$) were calculated from the output characteristics shown in Fig. 4(a) using a difference method (Fig. 4(b)), where $I_D$ is the drain current and $V_D$ is the drain voltage. For the superlinear curves, the difference calculated by subtracting the zero-bias conductance from the $dI_D/dV_D$ value at 0.5 V should be positive, while it should be negative for the sublinear curves. This difference, $\Delta(dI_D/dV_D)$, and its temperature dependence are plotted against $V_G-V_{NP}$ in Fig. 4(c). As expected for the hole-doping contacts (Fig. 1(b)), the degree of nonlinearity is higher in the positive gate region. Furthermore, the $\Delta(dI_D/dV_D)$ values in the positive gate region show a temperature dependence, where the superlinearity decreases as the temperature increases. The superlinearity appears when the inter-band transport is the dominant contribution to the total device resistance. Thus, the decrease in superlinearity indicates that the inter-band transport becomes more efficient and does not limit the current transport, which suggests a thermally activated nature of the inter-band transport. The thermal activation transport was confirmed by making an Arrhenius plot of $I_D$ as shown in Fig. 4(d). The activation energy for $I_D$ at $V_D = 0.1$ V was determined to be 1.4 and 3.3 meV for $V_G = V_{NP} - 40$ and $V_{NP} + 40$, respectively. The activation energy is higher in the positive gate region where the inter-band transport occurs, as schematically shown in Fig. 1(b).

**4. Discussion**

A finite bandgap should be introduced in the BLG underneath the metal contact because of the interfacial CT, and thus, $R_C$ may also be affected by an energy barrier for charge carrier injection from the metal electrode into the semiconductor, i.e. a Schottky barrier. The bandgap introduced in BLG is dependent on the electrical displacement fields ($\overline{D}$ in Ref. 5). Figure 5(a) shows the relationship between the displacement field and the Fermi level shift $\Delta E_F$ under the metal contact. The displacement field corresponds one-to-one with the bandgap (see the self-consistent tight-binding calculation data in Fig. 4 of Ref. 5), and thus, the relationship between the $\Delta E_F$ under the contacts and the bandgap can be obtained as shown in Fig. 5(b). As clearly seen from this figure, the $\Delta E_F$ under the metal contacts is always larger than the bandgap. For example, the bandgap under the metal contacts in the BLG FET without the NDT layer (Fig. 2(b)) is estimated to be 0.04 eV, and is considerably smaller than the $\Delta E_F$ obtained here (0.14 eV). Thus, the BLG underneath the metal contact is always degenerately doped, and the Schottky barrier at the metal-BLG contacts can be excluded as the origin of $R_C$.

As depicted in Fig. 6(a), several mechanisms can be considered for the inter-band transport: direct tunnelling, thermally activated tunnelling, and variable range hopping via in-gap states. The tunnelling mechanisms are understood by field-assisted (Fowler-Nordheim) tunnelling, and thus, the carrier transport should be enhanced in high $V_D$ regions and cause the superlinear $I_D$–$V_D$ characteristics shown in Fig. 4(a). Among these mechanisms, direct tunnelling can be excluded because the inter-band transport was shown to be thermally activated in Fig. 4(c). The direct tunnelling width can be estimated to be very long (roughly 60 nm), as shown in the potential profile in Fig. 6(b), indicating that the direct tunnelling process is unlikely to occur. In this profile, the interfacial region between the contact and the gated channel is simplified to be linear, and measures 0.5 µm in length, which is typical for metal contacts on single layer graphene [35]; the estimation was performed based on the under-contact $\Delta E_F$ of 0.14 eV of hole doping and $V_G$ of $V_{NP}$ + 40 V. The thermally activated tunnelling width is dependent on the thermal excitation energy. Figure 6(c) shows the thermally activated tunnelling width as a function of the thermal excitation energy. Tunnel conduction becomes efficient if the tunnelling barrier width is typically less than 5 nm, and a thermal activation of roughly 27 meV is found to be necessary to induce the thermally activated tunnelling. However, the activation energy estimated from Fig. 4(d) was one order of magnitude lower than 27 meV. Therefore, the only remaining mechanism, i.e. variable range hopping via in-gap states is the most likely mechanism of the inter-band transport. The possible origins of the in-gap states are the border trap and the local breakdown of the AB stacking [36], though it is difficult to determine the main origin at present.

The insertion of the NDT interfacial layer was found to affect the $R_C$, as shown in Fig. 3(b).

First, the determined $R_C$ values of the BLG FETs with and without the NDT layer are comparable to each other in the highly negative gate region, where the inter-band transport is not included, as schematically shown in Fig. 1(b). This indicates that a decrease in $R_C$ by the CT-induced DOS increase was compensated by other effects, most likely by a tunnel resistance of the NDT interfacial layer. Conductance of a tunnel junction is proportional to $\exp(-\beta l)$, where $\beta$ is the attenuation constant and $l$ is the junction length; $\beta$ is ca. 0.3 Å$^{-1}$ for oligophenyl molecular junctions [37], and the $l$ of the NDT layer is ~0.6 nm. Therefore, the conductance of the system should be reduced to 84%, corresponding to a resistance increase to 119%. The similar $R_C$ values of both systems with and without the NDT layer suggest that the resistance increase by the additional tunnel resistance was compensated by the decrease in $R_C$ by the CT-induced DOS increase, which is also ~84%. From the considerations above, the strategy for reducing $R_C$ by the CT-induced DOS increase is confirmed to operate properly in the negative gate region, but in the case of the NDT layer insertion, the total $R_C$ was not reduced due to the additional tunnel resistance. Thus, alternative device structures without inserting the molecular layer into the metal–BLG contacts should be employed to exclude the increase in $R_C$ because of the tunnel resistance across the molecular layer (Fig. 7).

Second, the determined $R_C$ values of the BLG FETs with the NDT layer are considerably higher than those without the NDT layer around the charge neutrality point. This feature might be ascribable to the device fabrication process. In the electron-beam lithography process, forward scattering of incident electrons is known to result in resist undercuts. Therefore, voids should remain after the electrode metal deposition. To the contrary, the NDT treatment process was conducted in the liquid phase, and should influence also the void parts. As a result, the part of the channel, which is close to the electrodes, should be affected by the NDT layer. NDT molecules on the channel is known to be an additional scattering source [23]. Around the charge neutrality point, the carrier concentration near the channel centre is small, and the high-resistive channel limits the carrier transport. On the other hand, the contact region limits the transport when the channel is highly doped by gating. Thus, the $R_C$ values around the charge neutrality point can be higher in the samples with the NDT layer.

Third, the determined $R_C$ values of the BLG FETs with the NDT layer are considerably higher than those without the NDT layer in the positive gate region. As discussed above, the effect of a decrease in $R_C$ because of the increase in the DOS is compensated by the increase in $R_C$ by the tunnel resistance. Thus, the higher $R_C$ can be understood by the effect of inter-band transport. Figure 6(d) compares the potential profiles with and without the NDT layer. The bandgap induced by the metal contact doping is larger with the NDT layer because of the higher hole concentration of BLG underneath the metal electrodes. In addition, the width of the region where the band bending occurs becomes longer because the potential difference between the contact and the gated channel is larger with the NDT layer. As a result, the path of the inter-band transport is estimated to be longer with the NDT layer, resulting in a higher $R_C$ in the positive gate region. The $R_C$ becomes more than double of that without the NDT layer. Since the $R_C$ decrease because of the DOS increase was found to remain 84%, the $R_C$ increase because of the inter-band transport is larger than the DOS-related decrease in the positive gate region.

The effects of the interfacial CT at metal contacts on BLG on $R_C$ is compiled in Table 1. In the present study, the charge carrier type doped from the metal electrode contacts was a hole, but the case for the electron-doping contacts is also included in this table. The result of the electron-doping of the contact is completely opposite that of the hole-doping of the contacts that have been discussed above. Therefore, for electron-doping of contacts, the inter-band transport occurs in the negative gate region where *pn* junctions are formed inside the channel, and $R_C$ reduction by the CT-induced DOS increase is effective in the positive gate region where no *pn* junction is formed. To fully exploit the $R_C$ reduction by the CT-induced DOS increase, it is necessary to use different contact doping carrier types based on the channel transport type: i.e. hole-doping contacts for the hole (*p*-type) channel and electron-doping contacts for the electron (*n*-type) channels.

## 4. Conclusions

In conclusion, the effects of source/drain electrode contacts in BLG FETs have been investigated by resolving contact resistance into its component elements. Contact doping produces two competing effects on the contact resistance: a decrease because of an increase in the DOS of the BLG underneath the metal contact, and an increase because of the inter-band transport via a bandgap generated by contact doping and gating. For the latter, a variable range hopping via in-gap states was proposed to be a possible mechanism of the inter-band transport. While the DOS-related effect affects the electrical characteristics of BLG FETs over the entire $V_G$ range, the inter-band transport occurs only when the charge carrier type of the gated channel is opposite to the contact doping carrier type. The amount of decrease in $R_C$ because of the DOS-related effect was found to be smaller than that of the increase in $R_C$ because of the inter-band transport. Therefore, the strategy for $R_C$ reduction by the CT-induced increase in the DOS is effective only for a single channel transport type ($n$- or $p$-type), depending on the contact doping carrier type. Furthermore, if the CT is induced by a molecular layer which is inserted into the metal–BLG interface, charge carrier tunnelling through the molecular insertion layer adds the tunnel resistance, leading to an increase in the contact resistance. Therefore, a molecule/BLG/electrode structure should be employed instead in order to prevent the additional resistance.

**Acknowledgments**

This work was supported in part by the Special Coordination Funds for Promoting Science and Technology from the Ministry of Education, Culture, Sports, Science and Technology of Japan; JSPS KAKENHI Grant Numbers JP26107531, JP16H00921 in Scientific Research on Innovative Areas "Science of Atomic Layers"; and a technology research grant from JFE 21st Century Foundation. The graphite crystal used in the present study was supplied by M. Murakami and M. Shiraishi.

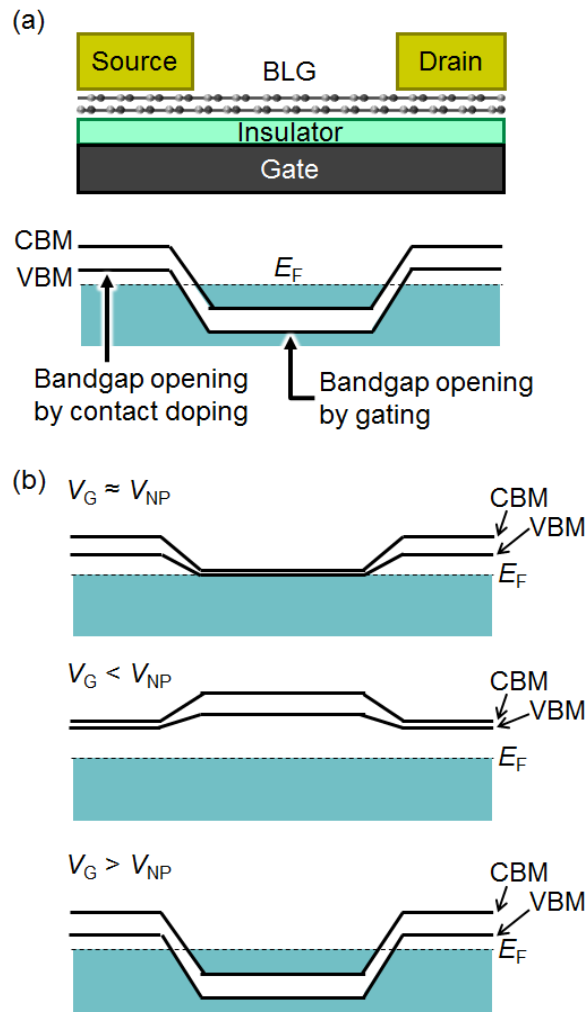

Figure 1. (a) Schematic diagram of the structure of a BLG FET. There are two sources of bandgap formation. CBM is the conduction band minimum, and VBM is the valence band maximum. (b) Inter-band transport at two *pn* junctions formed adjacent to the metal contacts and its gate-voltage dependence. The case of *p*-doping (hole-doping) contacts is depicted; the inter-band transport appears only when the channel is electrostatically *n*-doped (electron-doped) by gating.

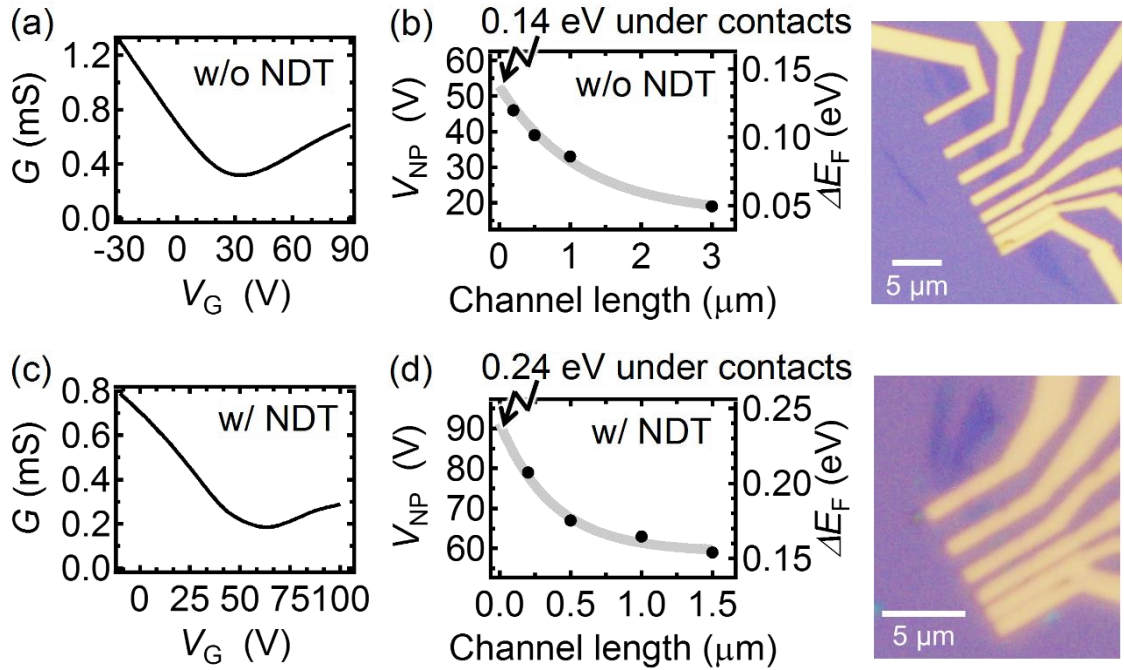

Figure 2. Electrical characteristics of as-fabricated BLG FETs without (a,b) and with (c,d) the NDT insertion layer. They were measured in air at room temperature under dark condition. (a,c) Transfer characteristics. The channel length $L$ and width $W$ are 1.0 and 2.6 μm for (a), and 1.0 and 1.3 μm for (c), respectively. Sheet conductances $G$ that can be calculated by $I_D L/(V_D W)$ are shown, where $I_D$ is the drain current and $V_D$ is the drain voltage. (b,d) Channel-length dependence of $V_{NP}$ displayed with a corresponding amount of Fermi level shift $\Delta E_F$ from the charge neutrality point. The right panels show an optical micrograph of the tested FETs.

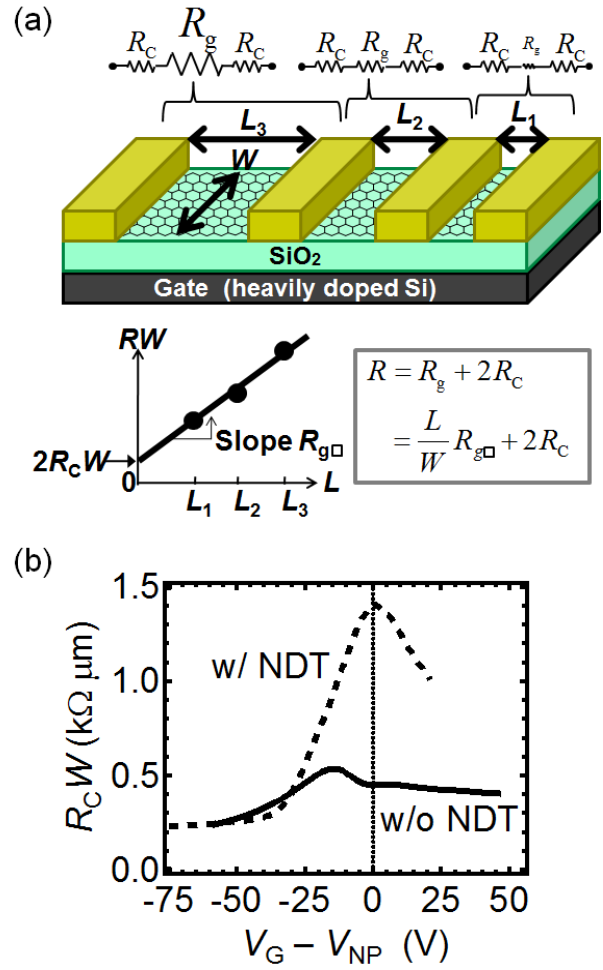

Figure 3. Extraction of contact resistance, $R_C$, from a channel-length dependence of transfer characteristics of the data in Fig. 2. (a) Extraction method using the TLM. $R_C$ can be extracted from the *y*-intercept of the total device resistance $R$ vs. the channel length $L$ plots. (b) Determined $R_C$ values normalized with the channel width, $W$.

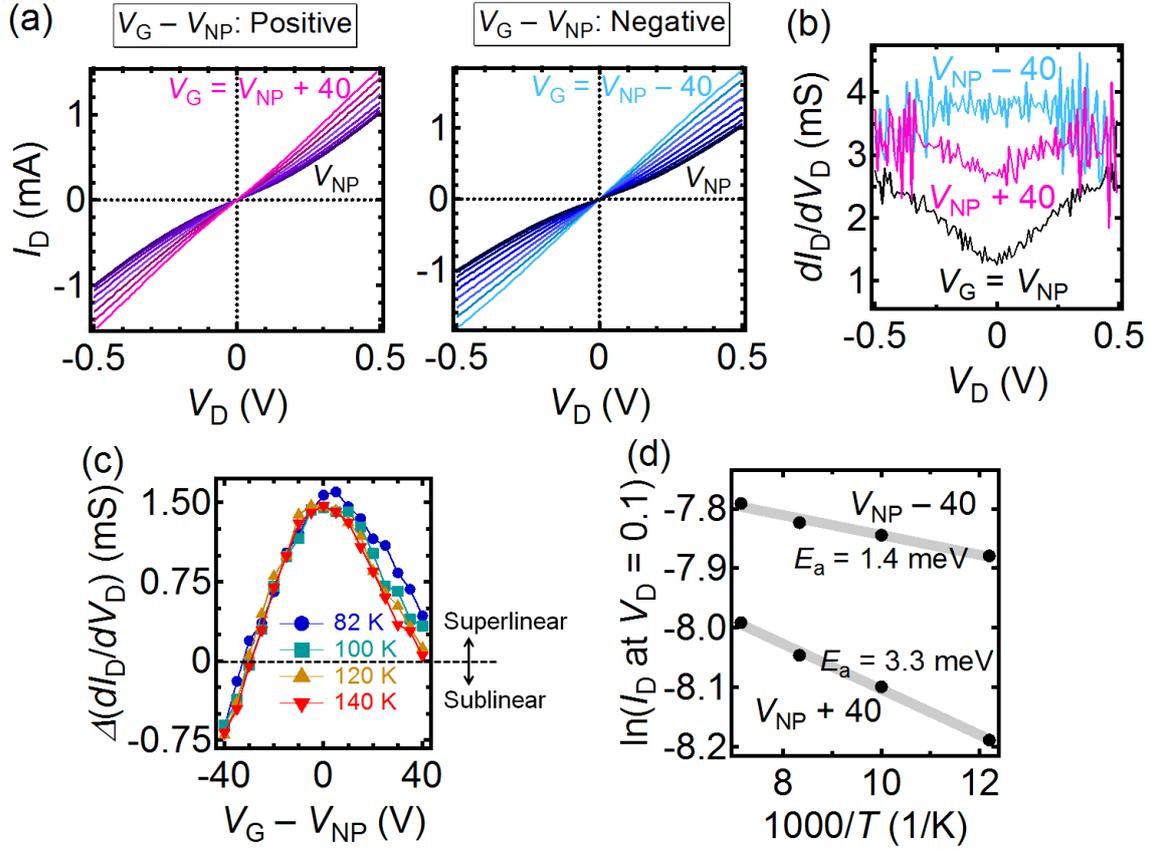

Figure 4. High-field output characteristics of a BLG FET without the NDT insertion layer. The measured FET is identical to the one for the 0.5-μm channel in Fig. 2(b), and the channel width is 3.0 μm. (a) Output characteristics measured at 82 K after annealing at 500 K for 3 h in a vacuum of $3 \times 10^{-3}$ Pa. The $V_G$ was swept from $V_{NP} - 40$ to $V_{NP} + 40$ V in 5 V steps. $I_D$ is the drain current, and $V_D$ is the drain voltage. (b) Differential conductance ($dI_D/dV_D$) calculated from (a) using a difference method. (c) Gate-voltage dependence of the degree of nonlinearity, $\Delta(dI_D/dV_D)$, which is defined as a difference calculated by subtracting the zero-bias conductance from the $dI_D/dV_D$ value at 0.5 V, and its temperature dependence. (d) Arrhenius plots of $I_D$ at $V_D = 0.1$ V for $V_G = V_{NP} - 40$ and $V_{NP} + 40$. The activation energy, $E_a$, is represented by the grey fitting lines.

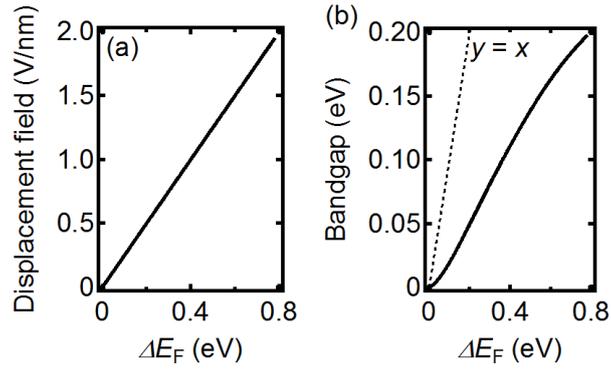

Figure 5. Bandgap of BLG underneath the metal contact. (a) Relationship between the electrical displacement field and the Fermi level shift under the metal contact. (b) Relationship between the Fermi level shift under the metal contact and the bandgap. The one-to-one correspondence between the displacement field and the bandgap is taken from the self-consistent tight-binding calculation data in Fig. 4 of Ref. 5.

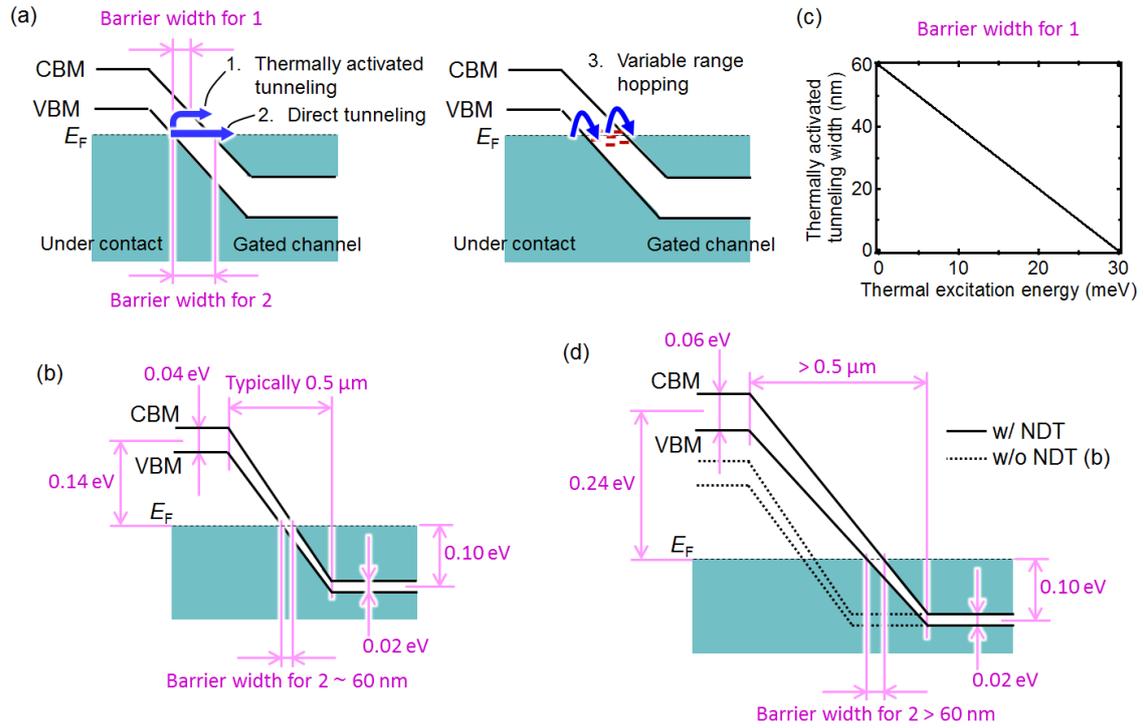

Figure 6. (a) Possible mechanisms of the inter-band transport. (b) Simplified potential profile that mimics the BLG FET without the NDT layer. The direct tunnelling width can be estimated to be 60 nm based on the under-contact $\Delta E_F$ of 0.14 eV of hole doping and $V_G$ of $V_{NP} + 40$ V. (c) Thermally activated tunnelling width as a function of the thermal excitation energy, which is estimated from the simplified potential profile in (b). (d) Simplified potential profiles that mimic the BLG FETs with and without the NDT layer. The inter-band transport length is longer with a higher contact doping, i.e. with the NDT layer.

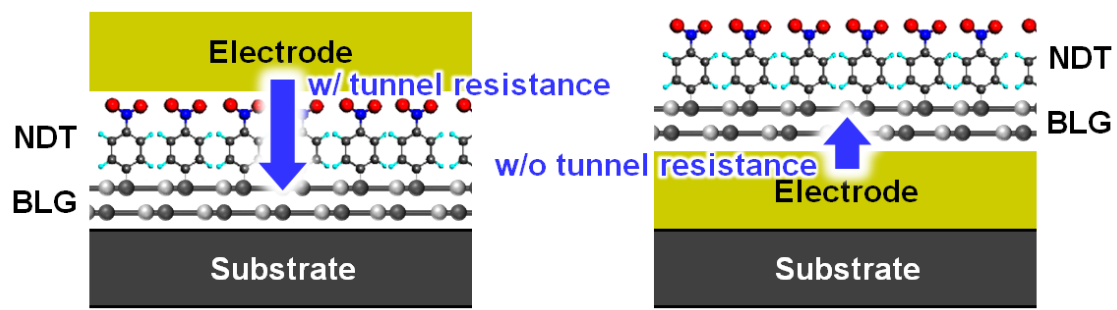

Figure 7. Additional tunnel resistance because of the molecular insertion layer (left panel). This resistance can be prevented by employing the molecule/BLG/electrode structure (right panel).

Table 1. Effects of the interfacial charge transfer at metal contacts on BLG on the contact resistance $R_C$.

| Contact doping | Mechanism | Change in $R_C$ | |
|---|---|---|---|
| | | $V_G < V_{NP}$ | $V_G > V_{NP}$ |
| Hole | Inter-band transport | (Not applicable) | Increase |
| | DOS increase | Decrease | Decrease, but smaller than the inter-band transport |
| Electron | Inter-band transport | Increase | (Not applicable) |
| | DOS increase | Decrease, but smaller than the inter-band transport | Decrease |

N, Żaba T, Cyganik P, Aspuru-Guzik A, and Whitesides G M 2016 *J. Phys. Chem. C* **120** 11331